\documentclass[preprint,12pt]{elsarticle}

\usepackage{graphicx}
\usepackage{amssymb}
\usepackage{multirow}
\usepackage{algorithm}
\usepackage{algpseudocode}
\usepackage{diagbox}
\usepackage{physics}
\usepackage{bm}
\usepackage{tikz}
\usepackage{comment}
\usepackage{graphicx}
\usepackage{subfig}
\usepackage{setspace}
\usepackage{float}

\usepackage{lineno}

\usepackage{breqn}
\usepackage{amsmath}

\DeclareMathOperator*{\argmin}{arg\,min}

\journal{Renewable and Sustainable Energy Reviews}

\begin{document}

\begin{frontmatter}


\title{Stochastic Pre-Event Preparation for Enhancing Resilience of Distribution Systems with High DER Penetration}




\author[mymainaddress]{Qianzhi Zhang\corref{mycorrespondingauthor}}
\cortext[mycorrespondingauthor]{Corresponding author}
\ead{qianzhi@iastate.edu}

\author[mymainaddress]{Zhaoyu Wang}

\author[mysecondaryaddress]{Shanshan Ma}

\author[mythirdaddress]{Anmar Arif}

\address[mymainaddress]{Department of Electrical and Computer Engineering, Iowa State University, Ames, IA 50011, USA}
\address[mysecondaryaddress]{School of Electrical, Computer and Energy Engineering, Arizona State University, Tempe, AZ 85287, USA}
\address[mythirdaddress]{Department of Electrical Engineering, King Saud University, Riyadh 11451, Saudi Arabia}

\begin{abstract}
This paper proposes a stochastic optimal preparation and resource allocation method for upcoming extreme weather events in distribution systems, which can assist utilities to achieve faster and more efficient post-event restoration. With the objective of maximizing served load and minimizing operation cost, this paper develops a two-stage stochastic mixed-integer linear programming (SMILP) model. The first-stage determines the optimal positions and numbers of mobile resources, fuel resources, and labor resources. The second-stage considers network operational constraints and repair crew scheduling constraints. The proposed stochastic pre-event preparation model is solved by a scenario decomposition method, Progressive Hedging (PH), to ease the computational complexity introduced by a large number of scenarios. Furthermore, to show the impact of solar photovoltaic (PV) generation on system resilience, we consider three types of PV systems during power outage and compare the resilience improvements with different PV penetration levels. Numerical results from simulations on a large-scale (more than 10,000 nodes) distribution feeder have been used to validate the scalability and effectiveness of the proposed method.
\end{abstract}

\begin{keyword}
Pre-event preparation \sep progressive hedging \sep PV systems \sep resource allocation \sep two-stage stochastic model
\end{keyword}

\end{frontmatter}


\section{Introduction}\label{sec:Intro}
Extreme weather events have brought significant damage to power grid infrastructure and caused 50\%-60\% of power outages in the U.S. \cite{salman2015evaluating}. Among those outages, around 90\% of them were due to failures in distribution systems \cite{WH_report}. After severe weather events, the major challenge for utilities is the shortage of various resources to repair damage and restore power supply. Pre-event resource allocation is one of the most effective ways to mitigate extreme events' impacts on power distribution system. It can allocate appropriate amounts of flexible resources to optimal positions before the extreme events. These flexible resources include emergency power supply resources, equipment resources and labor resources. Therefore, pre-event preparation enables faster and more efficient post-event restoration of the power gird. 

There are exist studies that have investigated resource allocation problems for the resilience enhancement of electric distribution systems. In \cite{pre_3,pre_4,pre_7}, proactive resource management in microgrids and proactive operation strategies in distribution systems are considered to enhance system resilience during extreme events. In \cite{pre_9}, the number and location of depots are determined at the pre-disturbance stage to manage the available resources. In \cite{B_1}, repair crews are pre-allocated to depots and integrated with restoration process to enhance the resilience of electric distribution systems. In \cite{B_3}, a two-stage stochastic model is developed to select staging locations and allocate repair crews for disaster preparation, while considering distribution system operation and crew routing constraints. In \cite{pre_1}, the authors developed a stochastic model for optimizing proactive operation actions. The study 
optimized the topology of the network and position of crews for upcoming disturbances. In \cite{pre_5} and \cite{pre_6}, a two-stage framework is developed to position mobile emergency generators (MEGs) for pre- and post-disasters. Mobile energy storage devices (MESs) are investigated in \cite{pre_2} and \cite{pre_10} for resilience enhancement of power distribution systems. However, there remain limitations in the above studies on pre-event preparation and resource allocation. These limitations are described in the following:

(1) \textit{Pre-allocation of various flexible resources:} In practice, pre-event preparation includes allocation of various flexible resources, such as MEGs, MESs, fuel resources for diesel generators, and repair crews. The optimal allocation of those flexible resources can help utilities to achieve faster and more efficient post-event power restoration. However, previous studies mainly focused on allocating specific flexible resources, rather than formulating a complete optimization problem to pre-allocate various flexible resources together.

(2) \textit{Impacts of solar PV power on system resilience:} Due to intermittent characteristic of traditional distributed energy resources (DERs), such as solar power, PV systems are not considered as a reliable resilient solution. However, the distributed nature of PV power can contribute to a more resilient power system. In practice, PV systems can be coupled with energy storage technology, to enable continues operation during outages \cite{NREL_report}. However, different types of PV systems are ignored in most existing research.    

(3) \textit{Scalability of the solution algorithm:} On one side, the stochastic pre-event preparation model may suffer from computational inefficiency due to a large number of scenarios; on the other side, a limited number of scenarios may influence the stability and quality of the solutions. Therefore, the trade-off between computation time and solution accuracy needs to be studied for stochastic pre-event preparation methods. In addition, a large-scale system is needed to verify the scalability of solution algorithms.    

To address these challenges, we propose a two-stage stochastic mixed integer linear program (SMILP) for pre-event preparation with pre-allocation of mobile resources, fuel resources and labour resources. Furthermore, the proposed pre-event preparation model considers different types of PV systems and facilitates the benefits of leveraging high PV penetration for improving the resilience of distribution grids. In this paper, resilience improvement is quantified by the increased served load and reduced outage duration. To deal with the massive computation burden, the proposed two-stage stochastic pre-event preparation problem is solved by a scenario decomposition method, Progressive Hedging (PH) \cite{T2_7}, while maintaining the accuracy and stability of the solution \cite{T2_8}. Also, the quality of the solution is validated by a multiple replication procedure (MRP). The main contribution of this paper is three-folded:
\begin{itemize}
\item We propose a two-stage SMILP model for pre-event preparation, where the first-stage allocates MEGs, MESs, fuel, and repair crews, while the second-stage considers distribution system operation and repair crew scheduling constraints.
\item The proposed model considers three types of PV systems. We also demonstrate the improvement of resilience and the reduction of outage duration with different PV penetration levels.  
\item The proposed solution algorithm is tested through a solution validation method to show its quality. In addition, a large-scale system, consisting of more than 10,000 nodes, is used to verify the scalability of the proposed pre-event preparation model.
\end{itemize}

The remainder of the paper is organized as follows: Section \ref{sec:Two_stage} describes the proposed two-stage SMILP for pre-event preparation and resource allocation. Section \ref{sec:Alg} presents the PH solution algorithm, convergence analysis and solution validation. Simulation results and conclusions are given in Section \ref{sec:Results} and Section \ref{sec:Con}, respectively.

\section{Two-stage Stochastic Pre-event Preparation Model}\label{sec:Two_stage}
\begin{figure}[h]
\centering\includegraphics[width=0.85\linewidth]{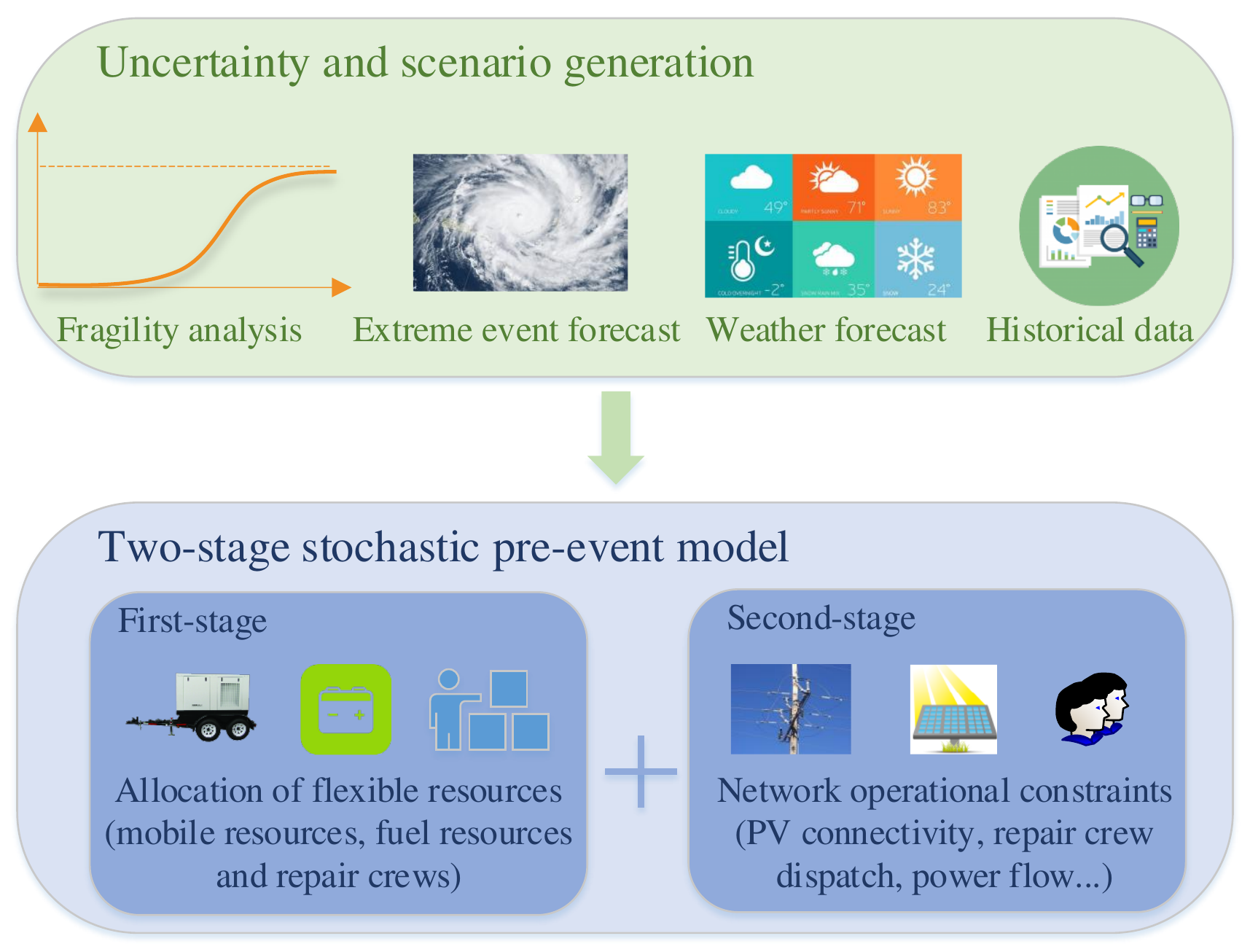}
\caption{The proposed two-stage stochastic pre-event model.}
\label{framework}
\end{figure}

The general framework of the proposed two-stage stochastic pre-event preparation model is shown in Figure \ref{framework}. 
Damage scenarios for extreme weather events are generated based on: (1) identification of extreme weather events, such as flood, hurricane and winter storm; (2) extreme weather event data and metric; (3) fragility model of test systems, which describes the behavior of components under extreme weather events; (4) damage status of components in test system subject to specific extreme weather events. To approximate the impact of extreme weather events to grid infrastructures, damage scenarios can be generated by mapping the weather data set to failure probability of grid infrastructures. Adopted from \cite{scen_ma}, the failure probability of an overhead line being damaged by hurricane can be expressed as follows:
\begin{equation}\label{scen_1}
p_{l,ij}(w(t))=1-\prod_{k=1}^{m}\Big(1-p_{l_k}(w(t))\Big)\prod_{k=1}^{n}\Big(1-p_{fc,k}(w(t))\Big)
\end{equation} 
where $p_{l,ij}(w(t))$ is the failure probability of the overhead line $ij$ with wind speed $w(t)$. $p_{l_k}(w(t))$ is defined as the conditional failure probability of pole $k$ at line $ij$ as a log-normal cumulative distribution function (CDF) of the wind speed $w(t)$, which is expressed in equation \eqref{scen_2}. $m$ and $n$ are the number of distribution poles supporting line $ij$ and the number of conductor wires between two adjacent poles at line $ij$, respectively. In equation \eqref{scen_3}, $p_{fc,k}(w(t))$ represents the failure probability of conductor $k$ between two poles.  
\begin{equation}\label{scen_2}
p_{l_k}(w(t))=\Phi\Big[\ln(\frac{w(t)/m_R}{\xi_R})\Big]
\end{equation} 
\begin{equation}\label{scen_3}
p_{fc,k}(w(t))=(1-p_u)\max\Big(p_{fw,k}(w(t)),\alpha p_{ftr,k}(w(t))\Big)
\end{equation} 
where $m_R$ and $\xi_R$ are the median capacity and the logarithmic standard deviation of intensity measurement, respectively. $p_{fw,k}(w(t))$ represents the direct wind-induced failure probability of conductor $k$ and $p_{ftr,k}(w(t))$ represents the fallen tree-induced failure probability of conductor $k$. $p_u$ is the probability that conductor $k$ is underground, which is more invulnerable to extreme weather events. $\alpha$ represents the average tree-induced damage probability of overhead conductors. More details of weather forecasting methodologies, line fragility models and scenario generation can be found in \cite{B_4}. 

As shown in Figure \ref{framework}, the proposed SMILP pre-event preparation model has two stages: (i) Flexible resources are allocated in the first-stage, including the optimal number and position of MEGs and MESs, allocation of available fuel to generators, and pre-position of repair crews to depots. (ii) The second-stage optimizes the operation of the distribution system and assign crews to the damaged components. Constraints in the second-stage includes unbalanced optimal power flow constraints, network reconfiguration and isolation constraints, and repair crew scheduling constraints. 

\subsection{Model Objective Function}
The objective function \eqref{obj_fun} is set to minimize operation costs and maximize load served. There are three cost related terms in the objective, cost of fuel $C^{\rm F}$, cost of switching operation $C^{\rm SW}$, and cost of load shedding $C_i^{\rm D}$. The objective is formulated as follows:
\begin{align}\label{obj_fun}
	\min\sum_{\forall s}Pr(s) &\Big(C^{\rm F}r^{\rm F}\sum_{\forall t}\sum_{\forall \phi}\sum_{\forall i}P^{\rm G}_{i,\phi,t,s}+C^{\rm SW}\sum_{\forall t}\sum_{\forall k \in\Omega_{\rm SW}}\gamma_{k,t,s}\nonumber\\
	&+\sum_{\forall t}\sum_{\forall \phi}\sum_{\forall i}C_i^{\rm D}(1-y_{i,t,s})d_{i,\phi,t}^{\rm p}\Big)    
\end{align}
where $Pr(s)$ is the probability of occurrence for scenario $s$, $r^{\rm F}$ is the rate of fuel consumption of a generator, and $P_{i,\phi,t,s}^{\rm G}$ is the active power output for fuel-based generator at bus $i$, phase $\phi$, time $t$, and scenario $s$. Binary variable $\gamma_{k,t,s}$ represents the status of each switch, if switch $k$ is operated at time $t$ on scenario $s$, then $\gamma_{k,t,s}=1$. The binary variable $y_{i,t,s}$ represents the status of load at bus $i$, time $t$, and scenario $s$. If the demand ($d_{i,\phi,t}^{\rm p}$) is served, then $y_{i,t,s}=1$.

\subsection{First-Stage Constraints}
The first-stage constraints revolve around pre-allocating four critical resources that will be utilized after an extreme event: (i) MEGs, (ii) MESs, (iii) fuel and (iv) repair crews. 

\subsubsection{Mobile Resources Allocation Constraints}
Mobile resources can be used to restore energy for isolated areas that are not damaged, and to restore critical customers. In addition, fuel management is critical after an extreme event to operate emergency generators. Distributing fuel after an extreme event maybe difficult due to road conditions. As for repair crews, pre-assigning them to different locations provides a faster and more organized response. The constraints for allocating the mobile resources are modeled as follows:
\begin{equation}\label{MEG}
	\sum_{\forall i\in\Omega_{\rm CN}}n_i^{\rm MEG}=N^{\rm MEG}
\end{equation}
\begin{equation}\label{MES}
	\sum_{\forall i\in\Omega_{\rm CN}}n_i^{\rm MES}=N^{\rm MES}
\end{equation}
\begin{equation}\label{MEG_MES}
	n_i^{\rm MEG}+n_i^{\rm MES}\leq N_i^{\rm MU},\forall i\in\Omega_{\rm CN}
\end{equation}
where binary variables $n_i^{\rm MEG}$ and $n_i^{\rm MES}$ equal 1 if an MEG and MES are allocated to bus $i$, respectively. The set $\Omega_{\rm CN}$ represents the set of candidate buses for MEGs and MESs. Constraints \eqref{MEG} and \eqref{MES} indicates that the number of installed MEGs and MESs are equal to the number of available devices ($N^{\rm MEG}$ and $N^{\rm MES}$). We assume that each bus can only have a limited number of mobile units $N_i^{\rm MU}$, which is enforced by \eqref{MEG_MES}.  

\subsubsection{Fuel Resources Allocation Constraints}

Define the set $\Omega_{\rm G}=\Omega_{\rm EG}\cup\Omega_{\rm CN}$, where $\Omega_{\rm EG}$ is the set of buses that have fuel-based emergency generators. The fuel allocated to $\Omega_{\rm G}$ must be limited to the available amount of fuel. We model the fuel allocation constraints as follows:
\begin{equation}\label{fuel_1}
	\sum_{\forall i\in\Omega_G}n^{\rm Fuel}_i\leq N^{\rm Fuel}
\end{equation}
\begin{equation}\label{fuel_2}
	F_i^{\rm G}\leq n^{\rm Fuel}_i\leq F_i^{\rm max},\forall i\in\Omega_{\rm G}
\end{equation}

Constraint \eqref{fuel_1} limits the total amount of allocated fuel to the amount of fuel available ($N^{\rm{Fuel}}$), where $n^{\rm Fuel}_i$ is the amount of fuel allocated to the generator at bus $i$. Constraint \eqref{fuel_2} limits the amount of fuel on each site, where $F_i^{\rm G}$ is the amount of fuel already present for the generator at bus $i$, and $F_i^{\rm max}$ represents the maximum capacity of fuel at bus $i$.  

\subsubsection{Repair Crew Allocation Constraints}
In order to allocate the repair crews, we divide the network into different regions $\Omega_{\rm R}$. Each region will be assigned with different crews, who will conduct the repairs in that region. The repair crews are pre-positioned to the regions using constraints \eqref{crew_1} and \eqref{crew_2}, as follows:
\begin{equation}\label{crew_1}
	\sum_{\forall r\in\Omega_{\rm R}}n^{\rm Crew}_r= N^{\rm Crew}
\end{equation}
\begin{equation}\label{crew_2}
	N_r^{\rm Crew,min}\leq n^{\rm Crew}_r\leq N_r^{{\rm Crew,max}},\forall r\in\Omega_{\rm R}
\end{equation}
where $n^{\rm Crew}_r$ is the number of repair crews in region $r$ and $N^{\rm Crew}$ is the total number of crews. The number of repair crews is limited in each region, using $N_r^{\rm Crew,min}$ and $N_r^{{\rm Crew,max}}$, depending on the size and capacity of the staging locations.

\subsection{Second-Stage Constraints}
In the second-stage of the proposed pre-event preparation model, the constraints of PV systems and repair crew dispatch are mainly discussed. The model also considers unbalanced power flow constraints, voltage constraints and reconfiguration constraints \cite{T2_1,T2_2}. 

\subsubsection{PV System Constraints}
To fully investigate the impact of PV systems on system resilience, three types of PV systems \cite{T2_2} are considered in the second-stage, $\Omega_{\rm PV}=\Omega^{\rm G}_{\rm PV}\cup\Omega^{\rm H}_{\rm PV}\cup\Omega^{\rm C}_{\rm PV}$: (i) Type 1: on-grid (grid-following) PV ($\Omega^{\rm G}_{\rm PV}$), where during an outage, the PV is switched off. (ii) Type 2: hybrid on-grid/off-grid PV + energy storage system (ESS) ($\Omega^{\rm H}_{\rm PV}$), where the PV system operates on-grid in normal conditions, and off-grid during an outage. (iii) Type 3: grid-forming PV + ESS with grid-forming capability ($\Omega^{\rm C}_{\rm PV}$), this system can restore part of the network that is not damaged if the fault is isolated. The output power of the PV systems is determined using the following equations:

\begin{equation}\label{PV_1}
0\leq P^{\rm PV}_{i,\phi,t,s}\leq \frac{I_{r_{i,t,s}}}{1000 W/m^2}P_i^{\rm rate},\forall i\in\Omega_{\rm PV}/\Omega^{\rm G}_{\rm PV},\phi,t,s
\end{equation}
\begin{equation}\label{PV_2}
0\leq P^{\rm PV}_{i,\phi,t,s}\leq \chi_{i,t,s} \frac{I_{r_{i,t,s}}}{1000 W/m^2}P_i^{\rm rate},\forall i\in\Omega^{\rm G}_{\rm PV},\phi,t,s
\end{equation}
\begin{equation}\label{PV_3}
(P^{\rm PV}_{i,\phi,t,s})^2+(Q^{\rm PV}_{i,\phi,t,s})^2\leq (S^{\rm PV}_i)^2,\forall i\in\Omega_{\rm PV}/\Omega^{\rm G}_{\rm PV},\phi,t,s
\end{equation}
\begin{equation}\label{PV_4}
(P^{\rm PV}_{i,\phi,t,s})^2+(Q^{\rm PV}_{i,\phi,t,s})^2\leq \chi_{i,t,s}(S^{\rm PV}_i)^2,\forall i\in\Omega^{\rm G}_{\rm PV},\phi,t,s
\end{equation}

The active power output $P^{\rm PV}_{i,\phi,t,s}$ of a PV depends on the rating of the solar cell $P_i^{\rm rate}$ and the solar irradiance $I_{r_{i,t,s}}$ \cite{T2_6}. The generated output power from the PV can be determined in constraints \eqref{PV_1} and \eqref{PV_2}, respectively. The binary variable $\chi_{i,t,s}$ equals 1 if bus $i$ is energized at time $t$ and scenario $s$. Using advanced PV smart inverters \cite{QZ_CVR}, the PVs can provide reactive power support $Q^{\rm PV}_{i,\phi,t,s}$, which is constrained by the capacity $S^{\rm PV}_i$ in \eqref{PV_3} and \eqref{PV_4}. During an outage, on-grid PVs are disconnected and the on-site load is not served by the PVs, therefore, constraints (\ref{PV_2}) and (\ref{PV_4}) are multiplied by $\chi_{i,t,s}$. PV systems of types $\Omega^{\rm C}_{\rm PV}$ and $\Omega^{\rm H}_{\rm PV}$ can disconnect from the grid and serve the on-site load. 

\begin{figure}[h]
\centering\includegraphics[width=0.95\linewidth]{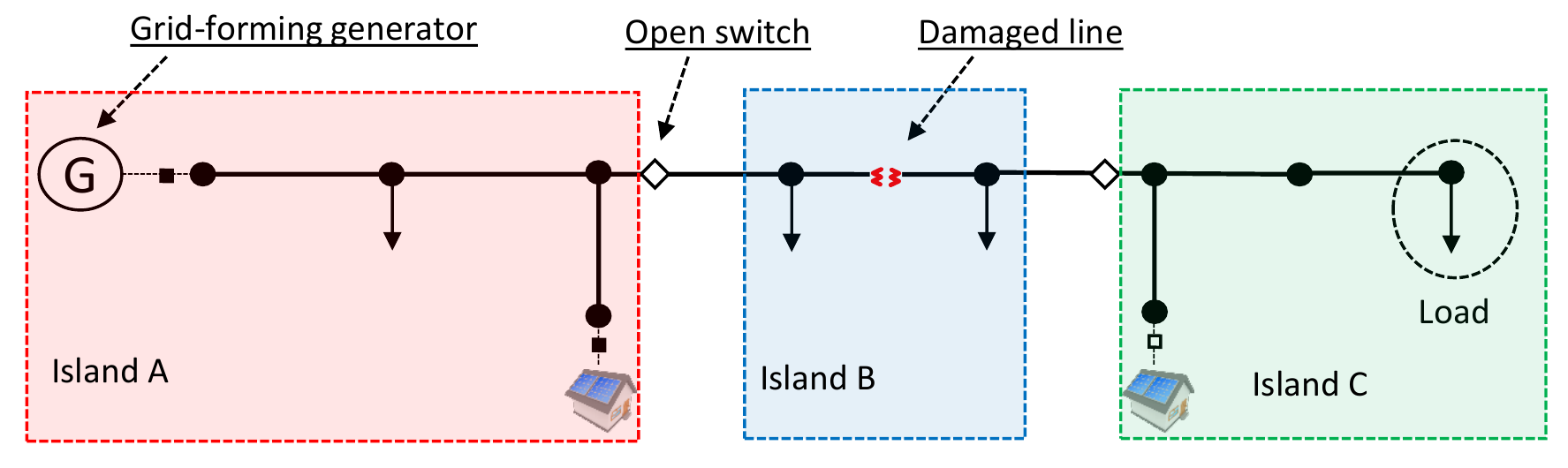}
\caption{A single line diagram of an example network with one damaged line.}
\label{damaged_line}
\end{figure}

An example network with damaged line is given in Figure \ref{damaged_line}, where the network is divided into three islands due to the damaged line. In this work, we assume that the network can be restored using the grid-forming sources in $\Omega^{\rm C}_{\rm PV}\cup\Omega_{\rm G}$. While PV system in types $\Omega^{\rm G}_{\rm PV}$ or $\Omega^{\rm H}_{\rm PV}$ can connect to the grid only after the PV bus is energized. Island A has a grid-forming generator, therefore, a microgrid is created and the PV system can participate. Island B must be isolated because of the damaged line. Island C does not have any grid-forming generators; hence, it will not be active and the grid-tied PV will be disconnected.

\begin{figure}[h]
\centering\includegraphics[width=0.95\linewidth]{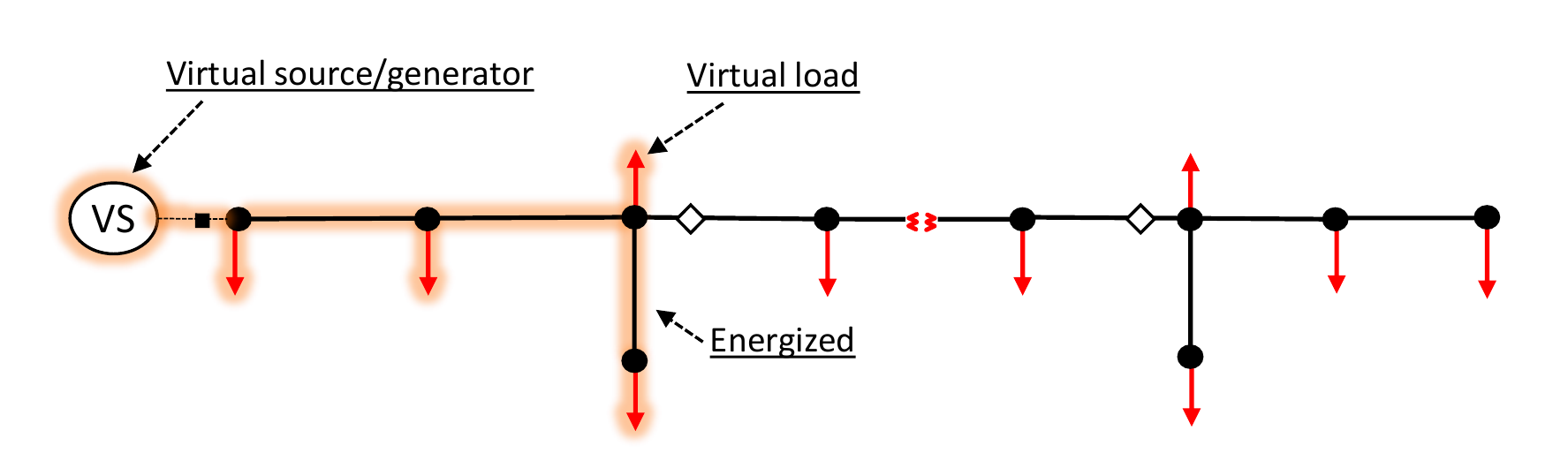}
\caption{A virtual network created for the example network in Figure \ref{damaged_line}.}
\label{virtual_network}
\end{figure}

To determine the connection status of the PV systems, we design a virtual network in parallel to the distribution network. The example network shown in Figure \ref{damaged_line} is transformed to a virtual network shown in Figure \ref{virtual_network}.
 A virtual network with virtual sources, loads, and flow is built to identify if an island can be energized by grid-forming generators. Each grid-forming generator is replaced by a virtual source with infinite capacity. Other power sources without grid-forming capability (e.g., grid-tied PVs) are removed. The actual loads are replaced by virtual loads with magnitude of 1. The virtual network scheme is modeled using constraints \eqref{virtual_1}-\eqref{virtual_5}.
 
 \begin{equation}\label{virtual_1}
\sum_{\forall j\in\Omega^{\rm C}_{\rm PV}\cup\Omega_{\rm G}}v_{j,t,s}^{\rm S}+\sum_{\forall k\in \Omega_{\rm K}(.,i)}v_{k,t,s}^{\rm f}= \chi_{i,t,s}+\sum_{\forall k\in \Omega_{\rm K}(i,.)}v_{k,t,s}^{\rm f},\forall i,t,s
\end{equation}
\begin{equation}\label{virtual_2}
	-(u_{k,t,s})M\leq v_{k,t,s}^{\rm f}\leq (u_{k,t,s})M,\forall k\in\Omega_{\rm K},t,s
\end{equation}
\begin{equation}\label{virtual_3}
	0\leq v_{k,t,s}^{\rm S}\leq (n_i^{\rm MEG}+n_i^{\rm MES})M,\forall i\in\Omega_{\rm CN},t,s
\end{equation}
\begin{equation}\label{virtual_4}
	\chi_{i,t,s}\geq y_{i,t,s},\forall i\in\Omega_{\rm N}/\{\Omega^{\rm C}_{\rm PV}\cup\Omega_{\rm PV}^{\rm H}\cup\Omega_{\rm G}\},t,s
\end{equation}
\begin{equation}\label{virtual_5}
	\chi_{i,t,s}+n_i^{\rm MEG}+n_i^{\rm MES}\geq y_{i,t,s},\forall i\in\Omega_{\rm CN},t,s
\end{equation}
 \noindent

A power-balance equation is added for each virtual bus, which means that if the virtual load at a bus is served, then that bus is energized. Therefore, for islands without grid-forming generators, all buses will be de-energized as the virtual loads in the island cannot be served. Constraint \eqref{virtual_1} is the node balance constraint for the virtual network. Virtual source $v^{\rm S}$ is connected to buses with power sources that have the capability to restore the system. The variable $v_k^{\rm f}$ represents the virtual flow on line $k$ and each bus is given a load of 1 that is multiplied by $\chi_i$. Therefore, $\chi_i=1$ (bus $i$ is energized) if the virtual load can be served by a virtual source and 0 (bus $i$ is de-energized) otherwise. The virtual flow is limited by \eqref{virtual_2}. The limits are multiplied by the status of the line ($u_{k,t,s}$) so that the virtual flow is 0 if a line is disconnected. The virtual source can be used only if a generator is installed, as enforced by \eqref{virtual_3}. Define $\Omega_{\rm N}$ as the set of all buses. If bus $i$ is de-energized, then the load must be shed \eqref{virtual_4}, unless bus $i$ has a local power source with disconnect switch. Constraint \eqref{virtual_5} is similar to \eqref{virtual_4} but with the presence of mobile sources.

\subsubsection{Repair Crews Constraints}
\begin{figure}[h]
\centering\includegraphics[width=0.95\linewidth]{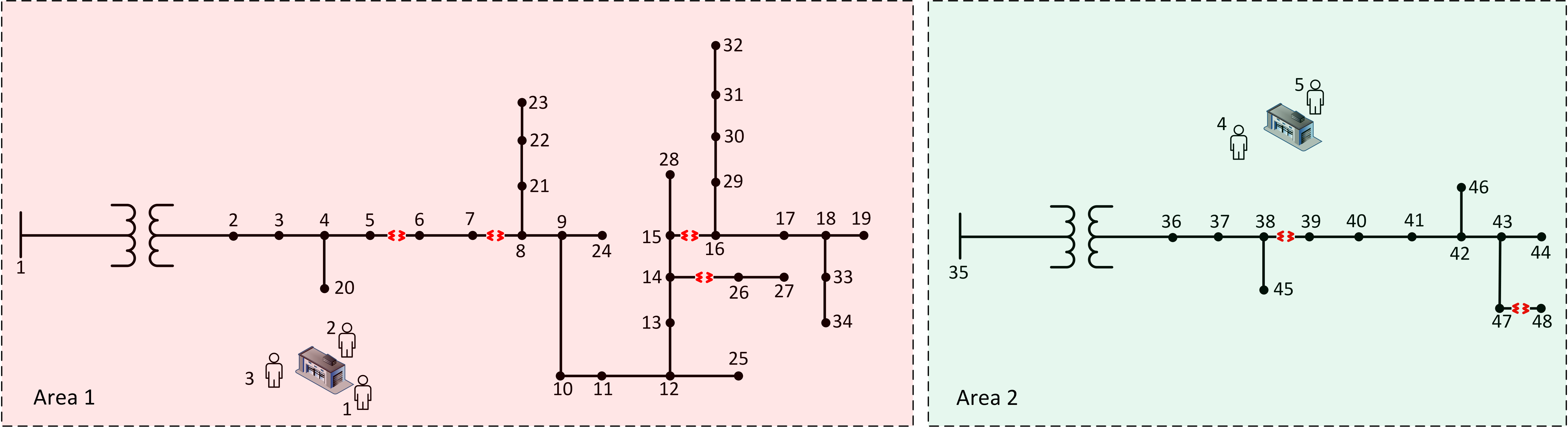}
\caption{A crew assignment example with 2 depots and 5 crews.}
\label{crews_example}
\end{figure}

In the second-stage, repair crews are assigned to damaged components that are in the area at which they are positioned. Note that the travel time is neglected in this study, as the travel distances between components in the same area is assumed to be small. An example for crew assignment is given in Figure \ref{crews_example}, where two working areas are assigned for the crews. In this example, four damaged lines in Area 1 will be repaired by crews 1-3, while crews 4 and 5 are responsible for the two damaged lines in Area 2. The repair crews constraints are formulated as follows:

\begin{equation}\label{crew_3}
\sum_{\forall k\in\Omega_{\rm DL(s)}}z_{k,t,s}\leq n^{\rm Crew}_r,\forall r,t,s
\end{equation}
\begin{equation}\label{crew_4}
\sum_{\forall t}z_{k,t,s}\leq T^r_{k,s},\forall k\in\Omega_{\rm DL(s)},s
\end{equation}
\begin{equation}\label{crew_5}
\frac{1}{T^r_{k,s}}\sum_{\tau=1}^{t-1}z_{k,\tau,s}-1+\epsilon \leq u_{k,t,s}\leq \frac{1}{T^r_{k,s}}\sum_{\tau=1}^{t-1}z_{k,\tau,s},\forall k\in\Omega_{\rm DL(s)},t,s
\end{equation}

\noindent

Define $z_{k,t,s}$ as a binary variable that equals 1 if line $k$ is being repaired at time $t$ on scenario $s$, and $\Omega_{\rm DL(s)}$ as the set of damaged lines on scenario $s$. Constraint \eqref{crew_3} limits the number of repairs being conducted in each area according to the number of crews $n^{\rm Crew}_r$ available. Constraint \eqref{crew_4} defines the repair time for each damaged line. The line status $u_{k,t,s}$ equals 0 until the repair process is conducted for $T_{k,s}^r$ time periods. Based on constraint \eqref{crew_5}, let $T_{k,s}^r=3$, $z_{k,t,s}=\{0,0,1,1,1,0,0\}$, then $u_{k,t,s}=\{0,0,0,0,0,1,1\}$. For example, when $t=6$ and $\epsilon=0.001$, then constraint \eqref{crew_5} becomes $0.668\leq u_{k,6,s}\leq 1$, therefore, $u_{k,6,s}=1$.

\subsubsection{Network Operational Constraints}
The next set of constraints are related to the operation of distribution systems. We consider unbalanced power flow equations, radiality constraints, fuel consumption, and energy storage constraints. The unbalanced distribution system constraints are given below:

\begin{align}\label{OPF_1}
\nonumber \sum_{b\in \Omega_{\rm K}(i,.)}P_{b,\phi,t,s}^{\rm K} - \sum_{k\in \Omega_{\rm K}(.,i)}&P_{k,\phi,t,s}^{\rm K} = P^{\rm G}_{i,\phi,t,s} + P^{\rm PV}_{i,\phi,t,s}\\
& + (P^{\rm Ch}_{i,\phi,t,s}-P^{\rm Dis}_{i,\phi,t,s}) - y_{i,t,s}d^P_{i,\phi,t},\forall i,\phi,t,s
\end{align}
\begin{align}\label{OPF_2}
\nonumber \sum_{b\in \Omega_{\rm K}(i,.)}Q_{b,\phi,t,s}^{\rm K} - \sum_{k\in \Omega_{\rm K}(.,i)}Q_{k,\phi,t,s}^{\rm K} &= Q^{\rm G}_{i,\phi,t,s} + Q^{\rm PV}_{i,\phi,t,s} \\
&+ Q^{\rm ESS}_{i,\phi,t,s} - y_{i,t,s}d^Q_{i,\phi,t},\forall i,\phi,t,s
\end{align}
\begin{equation}\label{OPF_3}
- u_{k,t,s}P_{k}^{\rm K,max}\leq P_{k,\phi,t,s}^{\rm K}\leq u_{k,t,s}P_{k}^{\rm K,max},\forall k \in \Omega_{\rm K},\phi,t,s
\end{equation}
\begin{equation}\label{OPF_4}
- u_{k,t,s}Q_{k}^{\rm K,max}\leq Q_{k,\phi,t,s}^{\rm K}\leq u_{k,t,s}Q_{k}^{\rm K,max},\forall k \in \Omega_{\rm K},\phi,t,s
\end{equation}
\begin{equation}\label{OPF_5a}
0 \leq P_{i,\phi,t,s}^{\rm G} \leq P_{i}^{\rm G,max} ,\forall i \in \Omega_{\rm EG},\phi,t,s
\end{equation}
\begin{equation}\label{OPF_6a}
0\leq Q_{i,\phi,t,s}^{\rm G} \leq Q_{i}^{\rm G,max} ,\forall i \in \Omega_{\rm EG},\phi,t,s
\end{equation}
\begin{equation}\label{OPF_5b}
0 \leq P_{i,\phi,t,s}^{\rm G} \leq n_{i}^{\rm{MEG}} P_{i}^{\rm G,max} ,\forall i \in \Omega_{\rm CN},\phi,t,s
\end{equation}
\begin{equation}\label{OPF_6b}
0\leq Q_{i,\phi,t,s}^{\rm G} \leq n_{i}^{\rm{MEG}} Q_{i}^{\rm G,max} ,\forall i \in \Omega_{\rm CN},\phi,t,s
\end{equation}
\begin{equation}\label{U_1}
\begin{split}
 U_{i,\phi,t,s}-U_{j,\phi,t,s} &\geq 2(\hat{R}_{ij}P_{ij,\phi,t,s}^{\rm K}+\hat{X}_{ij}Q_{ij,\phi,t,s}^{\rm K})\\
&+ (u_{k,t,s}+p_{ij,\phi}-2)M,\forall k,ij \in \Omega_{\rm K},\phi,t,s
\end{split}
\end{equation}
\begin{equation}\label{U_2}
\begin{split}
U_{i,\phi,t,s}-U_{j,\phi,t,s} &\leq 2(\hat{R}_{ij}P_{ij,\phi,t,s}^{\rm K}+\hat{X}_{ij}Q_{ij,\phi,t,s}^{\rm K})\\
&+ (2-u_{k,t,s}-p_{ij,\phi})M,\forall k,ij \in \Omega_{\rm K},\phi,t,s
\end{split}
\end{equation}
\begin{equation}\label{U_3}
\chi_{i,t,s}U^{\rm min}_{i} \leq U_{i,\phi,t,s} \leq \chi_{i,t,s}U^{\rm max}_{i},\forall i,\phi,t,s
\end{equation}
\begin{equation}\label{reconfigure_1}
\sum_{k \in \in\Omega_{\rm B(l)}}u_{k,t,s} \leq |\Omega_{\rm B(l)}|-1,\forall l\in\Omega_{\rm loop},t,s  
\end{equation} 
\noindent

Constraints \eqref{OPF_1} and \eqref{OPF_2} are the active and reactive nodal power balance constraints, where $P_{ij,\phi,t,s}^{\rm K}$ and $Q_{ij,\phi,t,s}^{\rm K}$ are the active and reactive line flows, and $P^{\rm G}_{i,\phi,t,s}$ and $Q^{\rm G}_{i,\phi,t,s}$ are the power outputs of the generators. The active charging/discharging and reactive power outputs of energy storage systems are denoted by $P^{\rm Ch}_{i,\phi,t,s}$, $P^{\rm Dis}_{i,\phi,t,s}$ and $Q^{\rm ESS}_{i,\phi,t,s}$. Constraints \eqref{OPF_3}-\eqref{OPF_4} represent the active and reactive power limits of the lines, where the limits ($P_{k}^{\rm K,max}$ and $Q_{k}^{\rm K,max}$) are multiplied by the line status binary variable $u_{k,t,s}$. Therefore, if a line is disconnected or damaged, power cannot flow through it. Constraints \eqref{OPF_5a}-\eqref{OPF_6a} limit the output of the generators to $P_{i}^{\rm G,max}$ and $Q_{i}^{\rm G,max}$. Similarly, we limit the output of the MEGs in \eqref{OPF_5b}-\eqref{OPF_6b} if an MEG is installed ($n_i^{\rm MEG} = 1$).

Constraints \eqref{U_1} and \eqref{U_2} calculate the voltage difference along line $k$ between bus $i$ and bus $j$, where $U_{i,\phi,t,s}$ is the square of voltage magnitude of bus $i$. We use the big-M method to relax constraints \eqref{U_1} and \eqref{U_2}, if lines are damaged or disconnected. $\hat{R}_{ij}$ and $\hat{X}_{ij}$ are the unbalanced three-phase resistance matrix and reactance matrix of line $ij$, which can be referred to \cite{QZ_CVR}. The vector $p_{ij,\phi}$ represents the phases of line $ij$. Constraint \eqref{U_3} guarantees that the voltage is limited within a specified region ($U^{\rm min}_{i}$ and $U^{\rm max}_{i}$), and is set to 0 if the bus is in an outage area. Constraint \eqref{reconfigure_1} can guarantee the radiality network during the network reconfiguration. In this paper, we assume that all the possible loops can be identified by depth-first search method. The set of loops are given by $\Omega_{\rm loop}$, and the set of switches in loop $l$ is given by $\Omega_{\rm B(l)}$. For each fuel-based generator, the total fuel consumption $F_{i,s}$ is limited by the available fuel resources $n^{\rm Fuel}_i$ in constraint \eqref{fuel_3}, as follows:
\begin{equation}
F_{i,s}=r^{\rm f}\sum_{\forall t}\sum_{\forall \phi}P^{\rm G}_{i,\phi,t,s} \leq n^{\rm Fuel}_i,\forall i\in\Omega_{\rm G},\phi,t,s\label{fuel_3}
\end{equation}

Next, we model the operation constraints for ESSs and MESs. The constraints include the change in state of charge (SOC), charging and discharging limits, and reactive power limits.  Let $\Omega_{\rm ES}$ be the set of buses with ESSs, and $\Omega_{\rm ESC} = \Omega_{\rm ES} \cup \Omega_{\rm CN}$. We can then define the energy storage constraints as follows:

\begin{equation}\label{ESS_1}
\begin{split}
E^{\rm SOC}_{i,t,s} = &E^{\rm SOC}_{i,t-1,s} +\\& \Delta t \frac{(\sum_{\forall \phi}P^{\rm Ch}_{i,\phi,t,s}\eta_{\rm Ch}- \sum_{\forall \phi}P^{\rm Dis}_{i,\phi,t,s}/\eta_{\rm Dis})}{E^{\rm Cap}_{i}},\forall i \in \Omega_{\rm ESC},\phi,t,s
\end{split}
\end{equation}
\begin{equation}\label{ESS_2}
E^{\rm SOC,min}_{i}\leq E^{\rm SOC}_{i,t,s}\leq E^{\rm SOC,max}_{i},\forall i \in \Omega_{\rm ESC},t,s
\end{equation}
\begin{equation}\label{ESS_3}
0\leq P^{\rm Ch}_{i,\phi,t,s}\leq h_{i,t,s}P^{\rm Ch,max}_{i},\forall i \in \Omega_{\rm ESC},\phi,t,s
\end{equation}
\begin{equation}\label{ESS_4}
0\leq P^{\rm Dis}_{i,\phi,t,s}\leq (1-h_{i,t,s})P^{\rm Dis,max}_{i},\forall i \in \Omega_{\rm ESC},\phi,t,s
\end{equation}
\begin{equation}\label{ESS_5}
-Q_{i}^{\rm ESS,max}\leq Q_{i,\phi,t,s}^{\rm ESS} \leq Q_{i}^{\rm ESS,max},\forall i \in \Omega_{\rm ES},\phi,t,s
\end{equation}
\begin{equation}\label{ESS_3b}
0\leq P^{\rm Ch}_{i,\phi,t,s}\leq n_i^{\rm MES}P^{\rm Ch,max}_{i},\forall i \in \Omega_{\rm CN},\phi,t,s
\end{equation}
\begin{equation}\label{ESS_4b}
0\leq P^{\rm Dis}_{i,\phi,t,s}\leq n_i^{\rm MES} P^{\rm Dis,max}_{i},\forall i \in \Omega_{\rm CN},\phi,t,s
\end{equation}
\begin{equation}\label{ESS_5b}
-n_i^{\rm MES} Q_{i}^{\rm ESS,max}\leq Q_{i,\phi,t,s}^{\rm ESS} \leq n_i^{\rm MES} Q_{i}^{\rm ESS,max},\forall i \in \Omega_{\rm CN},\phi,t,s
\end{equation}
\noindent

Constraint \eqref{ESS_1} determines the state of charge of ESSs ($E^{\rm SOC}_{i,t,s}$). $E^{\rm Cap}_{i}$ denotes the maximum capacity of the storage system. To ensure safe ESS operation, the SOC and charging ($P^{\rm Ch}_{i,\phi,t,s}$) and discharging ($P^{\rm Dis}_{i,\phi,t,s}$) power of ESSs are constrained as shown in \eqref{ESS_2}-\eqref{ESS_4}. Here, $E^{\rm SOC,min}_{i}$, $E^{\rm SOC,max}_{i}$, $P^{\rm Ch,max}_{i}$ and $P^{\rm Dis,max}_{i}$ define the permissible range of SOC, and maximum charging and discharging power, respectively. In constraints \eqref{ESS_3}-\eqref{ESS_4}, the binary variable $h_{i,t,s}$ indicates that ESSs cannot charge and discharge at the same time instant. The ESS charging/discharging efficiency are represented by $\eta_{\rm Ch}$/$\eta_{\rm Dis}$. The reactive power of ESS, $Q_{i,\phi,t,s}^{\rm ESS}$, is kept within maximum limit, $Q_{i}^{\rm ESS,max}$, through constraint \eqref{ESS_5}. For MES units, we add constraint \eqref{ESS_3b}-\eqref{ESS_4b} so that if $n_i^{\rm MES} = 0$, the output power is 0 at bus $i$. The same method is applied for the reactive power in \eqref{ESS_5b}.

\section{Solution Algorithm}\label{sec:Alg}
When the number of scenarios is finite, a two-stage stochastic problem can be modeled as a single-stage large linear programming model, where each constraint in the problem is duplicated for each realization of the random data. For problems where the number of realization is too large or infinite, the Monte Carlo sampling technique can be used to generate a manageable number of scenarios. In this work, we use the scenario decomposing method PH to solve the proposed two-stage stochastic pre-event preparation problem.  

\subsection{Two-stage Progressive Hedging Algorithm}
The proposed two-stage stochastic pre-event preparation problem \eqref{obj_fun}-\eqref{ESS_5b} can be compactly reformulated with an extensive form (EF) as follows:
\begin{equation}\label{solution_1}
\xi=\min_{x,y_s}a^Tx+\sum_{\forall s}Pr(s)b^T_s y_s
\end{equation}
\begin{equation}\label{solution_2}
\text{s.t.}\hspace{2mm} (x,y_s)\in Q_s,\forall s
\end{equation}
where $a$ and $b_s$ are vectors containing the coefficients associated with the compact first-stage variable $x$ and compact second-stage  variable $y_s$ in the objective \eqref{solution_1}, respectively. The constraint \eqref{solution_2} represents the subproblem constraints that ensure a feasible solution. The PH algorithm decomposes the extensive form into scenario-based subproblems, by relaxing the non-anticipativity of the first-stage variables. Hence, with the total number $S$ of scenarios, the proposed stochastic pre-event preparation problem is decomposed into $S$ subproblems. The proposed two-stage PH algorithm is presented in Algorithm 1. Define $\tau$ as iteration number, $\rho$ as a penalty factor and $\epsilon$ as a termination threshold. The PH algorithm starts by solving the subproblems with individual scenarios. Note that for an individual scenario, the two-stage model is reformulated to a single-level problem. In Step 4, the first-stage solution obtained from Step 2 is aggregated to obtain the expected value $\bar{x}$. Step 5 calculates the value of the multiplier $\eta_s$. In Step 8, the subproblems are solved, where each subproblem is augmented with a linear term proportional to the multiplier $\eta^{\tau-1}_s$ and a squared two norm term penalizing the difference of $x$ from $\bar{x}^{\tau-1}$. Steps 9-10 are similar as Steps 4-5. The algorithm terminates once all first-stage decisions $x_s$ converge to a common $\bar{x}$.
\begin{algorithm}[t]
\caption{The Two-Stage PH Algorithm}\label{alg1}
\begin{algorithmic}[1]
\State \hspace{0mm}{\bf Initialization}: Let $\tau:=0$.
\State For all $s\in S$, compute.
\State $x_s^{(\tau)}:=\argmin_x \{a^Tx+b^T_s y_s:(x,y_s)\in Q_s\}$.
\State $\bar{x}^{(\tau)}:=\sum_{\forall s\in S}Pr(s)x^{(\tau)}_s$.
\State $\eta^{(\tau)}_s:=\rho(x_s^{(\tau)}-\bar{x}^{(\tau)})$.
\State $\tau:=\tau+1$.
\State For all $s\in S$, compute.
\State $x_s^{(\tau)}:=\argmin_x \{a^Tx+b^T_s y_s+\eta^{(\tau-1)}_s x +\frac{\rho}{2}\|x_s^{(\tau)}-\bar{x}^{(\tau)}\|^2:(x,y_s)\in Q_s\}$.
\State $\bar{x}^{(\tau)}:=\sum_{\forall s\in S}Pr(S)x^{(\tau)}_s$.
\State $\eta^{(\tau)}_s:=\eta^{(\tau-1)}_s+\rho(x_s^{(\tau)}-\bar{x}^{(\tau)})$.
\If{$\sum_{\forall s\in S}Pr(s)\|x_s^{(\tau)}-\bar{x}^{(\tau)}\|\leq \varepsilon$}
\State Go to Step 5. 
\Else
\State terminate.
\EndIf
\end{algorithmic}
\end{algorithm}

\subsection{Convergence and Solution Validation}
As shown in Algorithm 1, the convergence metric $g^\tau$ of progressive hedging algorithm at each iteration $\tau$ is expressed as the deviation from the mean summed across all first-stage variables $x_s(\tau)$ and the average value of the first-stage variable $\bar{x}^\tau$ as follows:
\begin{equation}\label{con1}
g^\tau=\sum_{s\in S}Pr(s)\|x_s(\tau)-\bar{x}^\tau\|
\end{equation}

Numerical results for convergence analysis are given in case study section. In order to test the solution quality based on the limited generated damage scenarios, we follow the suggestion from \cite{T2_2} and apply MRP to test the stability and quality of the candidate solutions, as shown in Algorithm 2. MRP is to repeat the procedure of generating $S$ scenarios and solving the proposed model for $S$ times and construct the confidence interval (CI) for the optimality gap. The detailed steps in MRP is shown in Algorithm 2, where $\bar{G_n}(n_g)$ is the gap estimate and $s^2_G(n_g)$ is the sample variance.
\begin{algorithm}[t]
\caption{Multiple Replication Procedure}\label{alg2}
\begin{algorithmic}[1]
\State {\bf Input}: Value $\alpha\in (0,1)$ (e.g., $\alpha=0.05$), sample size $n$, replication size $n_g$ and a candidate solution $\hat{x}\in X$.
\State {\bf Output}: Approximate $(1-\alpha)$ as the level confidence interval on $\mu_{\hat{x}}$.
\State For $k=1,2,...,n_g$.
\State Sample i.i.d. observations $\zeta^{k_1},\zeta^{k_2},...,\zeta^{k_n}$ from the distribution of $\zeta$.
\State Solve ($\rm SP_n$) using $\zeta^{k_1},\zeta^{k_2},...,\zeta^{k_n}$ to obtain $x^{k*}_n$.
\State $G^k_n(\hat{x}):=n^{-1}\sum_{j=1}^n(f(\hat{x},\zeta^{kj})-f(x^{k*}_n,\zeta^{kj}))$.
\State $\bar{G_n}(n_g):=\frac{1}{n_g}\sum_{k=1}^{n_g}G^k_n(\hat{x})$.
\State $s^2_G(n_g):=\frac{1}{n_g-1}\sum_{k=1}^{n_g}(G^k_n(\hat{x})-\bar{G_n}(n_g))^2$.
\State $\epsilon:=t_{n_g-1,\alpha}S_G(n_g)/\sqrt{n_g}$.
\State Obtain one-sided CI on $[0,\bar{G_n}(n_g)+\epsilon_g]$.
\end{algorithmic}
\end{algorithm}

\section{Case Study}\label{sec:Results}
In this section, a large-scale system is used as a test case to verify the scalability and effectiveness of the two-stage stochastic pre-event resource allocation model. This large-scale system consists of 3 existing test systems, EPRI ckt5, ckt7 systems \cite{Opendss}, and IEEE 8500 bus system \cite{8500}, Following the suggestions from \cite{ma2020resilience}, the cost parameters in the simulation are $C^{\rm D}=14\$/kWh$, $C^{\rm SW}=8\$$, $C^{\rm F}=1\$/L$ and $r^{\rm F}=0.3L/kWh$. The stochastic models and algorithms are implemented using the PySP package in Pyomo \cite{T2_11}. IBM's CPLEX 12.6 mixed-integer solver is used to solve all subproblems. The experiments were performed on Iowa State University's Condo cluster, whose individual blades consist of two 2.6 GHz 8-Core Intel E5-2640 v3 processors and 128 GB of RAM.

\subsection{Pre-Event Preparation Results}
\begin{figure}[H]
\centering\includegraphics[width=0.95\linewidth]{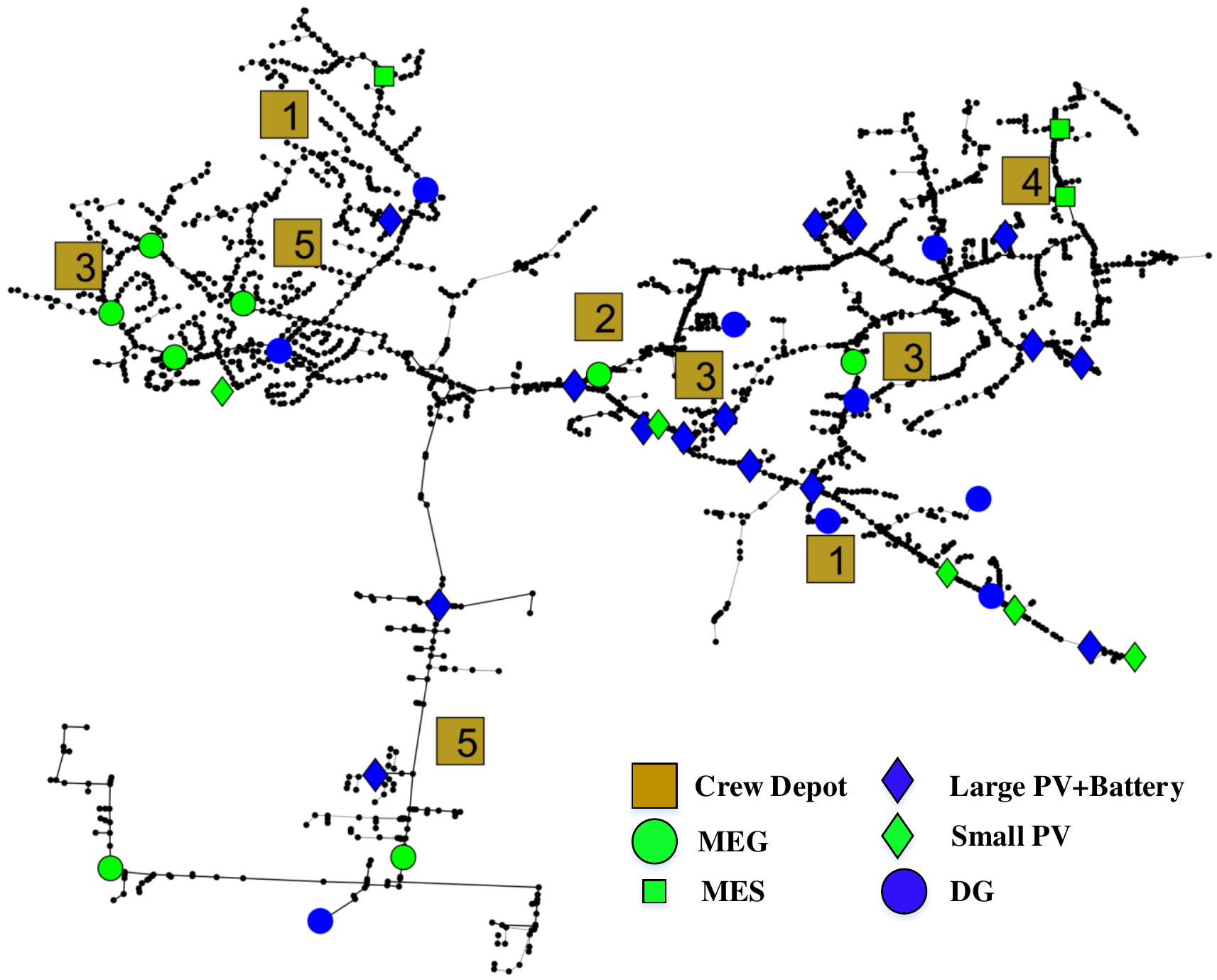}
\caption{Resource allocation of large-system with the proposed model.}
\label{large_res}
\end{figure}

In this case, we have included 9 depots that are hosting a total of 27 crews, 9 dispatchable DGs, 8 MEGs, 3 MESs, 123 switches, 5 small PVs, 15 large PVs, and 12 ESSs. The 9 DGs are rated at 300 kW and 250 kVAr. The 5 small PVs are rated at 11kW$\sim$22kW. The 15 large PVs are rated at 500 kW. The 12 ESSs are rated at 500 kW/ 3500 kWh. The pre-event preparation model of the large-scale system is solved in 10.2 hours with 10 damage scenarios. The first-stage decision variables (locations of MEGs, MESs and crews) are shown in Figure \ref{large_res}. 27 crews are allocated to 9 different depots. The value inside the crew depot in Figure \ref{large_res} represents the number of crews assigned to that depot. Areas with large number of crews indicates that the lines in the area have high damage probabilities. 

\begin{figure}[H]
\centering\includegraphics[width=0.85\linewidth]{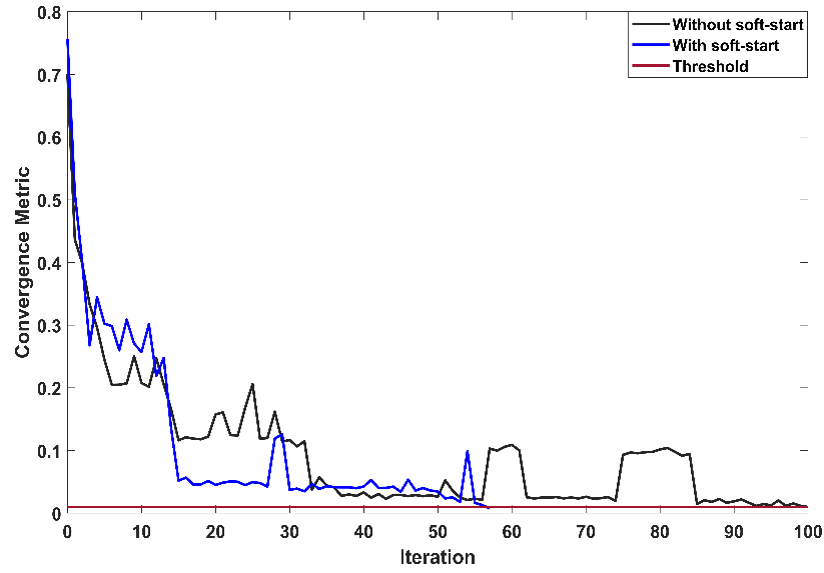}
\caption{The convergence metric comparison with and without soft-start solutions.}
\label{convergence}
\end{figure}

As discussed in Section 3.2, we use the convergence metric to evaluate the convergence speed of the proposed model. At the same time, we also compare the computational speed with and without a soft-start solution. Soft-start solution means that the previous computed solution in other instance will be used as the starting point. The comparison result is shown in Figure \ref{convergence}. If the convergence metric reaches the convergence threshold 0.01, the algorithm will stop and obtain the optimal solution. The instance with soft-start solution converges at 57 iteration and takes 10.2 hours. The case without soft-start solution converges after 100 iteration and takes 24.3 hours. To test the solution quality with MRP, based on the limited generated damage scenarios, the one-sided CI of the obtained solution is $[0,12.48\%]$. This small gap indicates that our solution is stable and of high quality.

\begin{figure}[H]
\centering\includegraphics[width=0.85\linewidth]{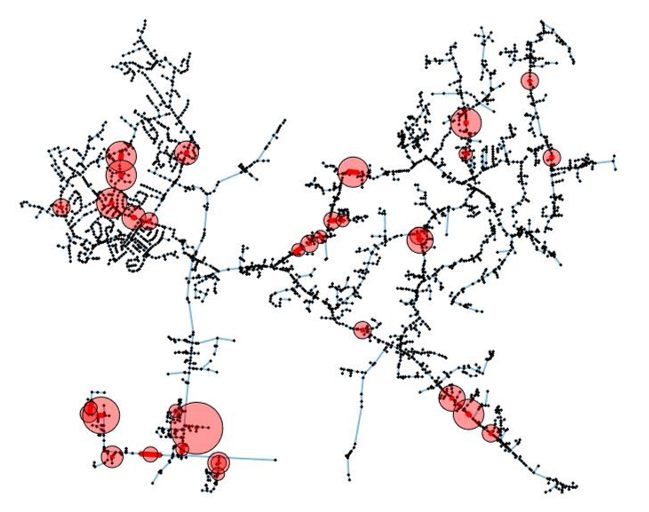}
\caption{Aggregated damaged areas.}
\label{Agg_damage}
\end{figure}

To evaluate the performance of the developed pre-event preparation model, the model is compared to a base model. The base case is generated by the following steps: (i) one MEG are prepositioned at the substations. (ii) Extra MEG are prepositioned at high-priority loads. (iii) PV and ESS are not considered. (iv) Fuel is allocated to the MEGs such that they can operate for at least 24 hours. (v) Crews are allocated evenly between depots. In this work, we calculate average outage duration by dividing the sum of outage durations for the loads with the number of loads. To compare the performance of the proposed model and the base model, we generate a random scenario and test the response of the system. The generated scenario has 103 damaged lines and they were aggregated to 34 damaged areas in Figure \ref{Agg_damage}. Each circle represents the repair time needed for the specific damaged area considering all the aggregated damaged lines. 

\begin{figure}[H]
\centering\includegraphics[width=0.85\linewidth]{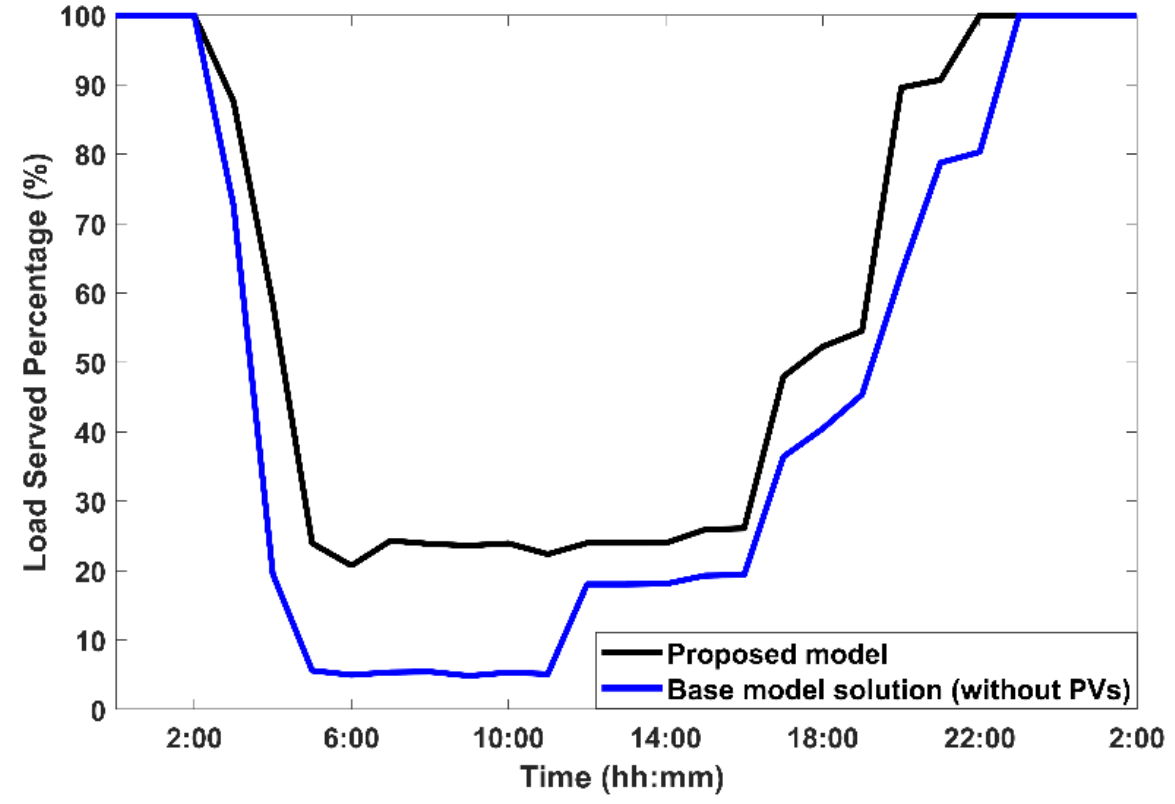}
\caption{Comparison between base model and proposed method.}
\label{large_com}
\end{figure}

The comparison between the base model and the proposed method is shown in Figure. \ref{large_com}. In the base model, the total restored energy is 231,422.38 kWh and the average outage duration is 14.69 hours. In the proposed method, the total restored energy is 291,727.48 kWh and the average outage duration is 11.28 hours. Therefore, approximately 20.67\% more loads are served by the proposed method and the outage duration decreased by 30.22\%.

\subsection{Impacts of Solar PV on System Resilience}
To show the advantages of the PV systems, we test the response of the system with the proposed method and different PV penetration levels. As discussed in Section 2.3.1, we consider three types of PV: (i) Type I PV, which represents residential PV panels and the rated capacity is assumed to be 6 kW; (ii) Type II PV, which represents mid-size PV systems and the rated capacity is assumed to be 48 kW; (iii) Type III PV, which represents large utility PV farm and the rated capacity is assumed to be 2000 kW. Based on the number of different types of PVs, we define 6 PV penetration levels as 9\%, 27\%, 45\%, 63\%, 81\%, and 99\%. The number of Type I, II and III PVs for each PV penetration levels is summarized in Table \ref{PV_pen}. To better collaborate the setting of PV penetration, the number of dispatchable DGs has been changed to 10 and the positions of those DGs have been changed accordingly. The rest of case settings keep the same. 
\begin{table}[H]
		\centering
		\renewcommand{\arraystretch}{1.3}		
		\caption{Number of different types of PV}
	    \label{PV_pen}
\begin{tabular}{lcccl}
\hline
\textbf{\begin{tabular}[c]{@{}l@{}}PV Penetration \\ Percentage\end{tabular}} & \multicolumn{1}{l}{\textbf{\begin{tabular}[c]{@{}l@{}}Type I \\ PV\end{tabular}}} & \multicolumn{1}{l}{\textbf{\begin{tabular}[c]{@{}l@{}}Type II \\ PV\end{tabular}}} & \multicolumn{1}{l}{\textbf{\begin{tabular}[c]{@{}l@{}}Type III \\ PV\end{tabular}}} 
\\\hline
\textbf{9\%}                                      & 8                                      & 1                                       & 1                                        &  \\
\textbf{27\%}                                    & 24                                     & 4                                       & 3                                        &  \\
\textbf{45\%}                                    & 40                                     & 7                                       & 5                                        &  \\
\textbf{63\%}                                    & 63                                     & 9                                       & 7                                        &  \\
\textbf{81\%}                                     & 72                                     & 12                                      & 9                                        &  \\
\textbf{99\%}                                     & 88                                     & 15                                      & 11  \\\hline                                          & 
\end{tabular}
\end{table}

\begin{figure}[H]
\centering\includegraphics[width=1.0\linewidth]{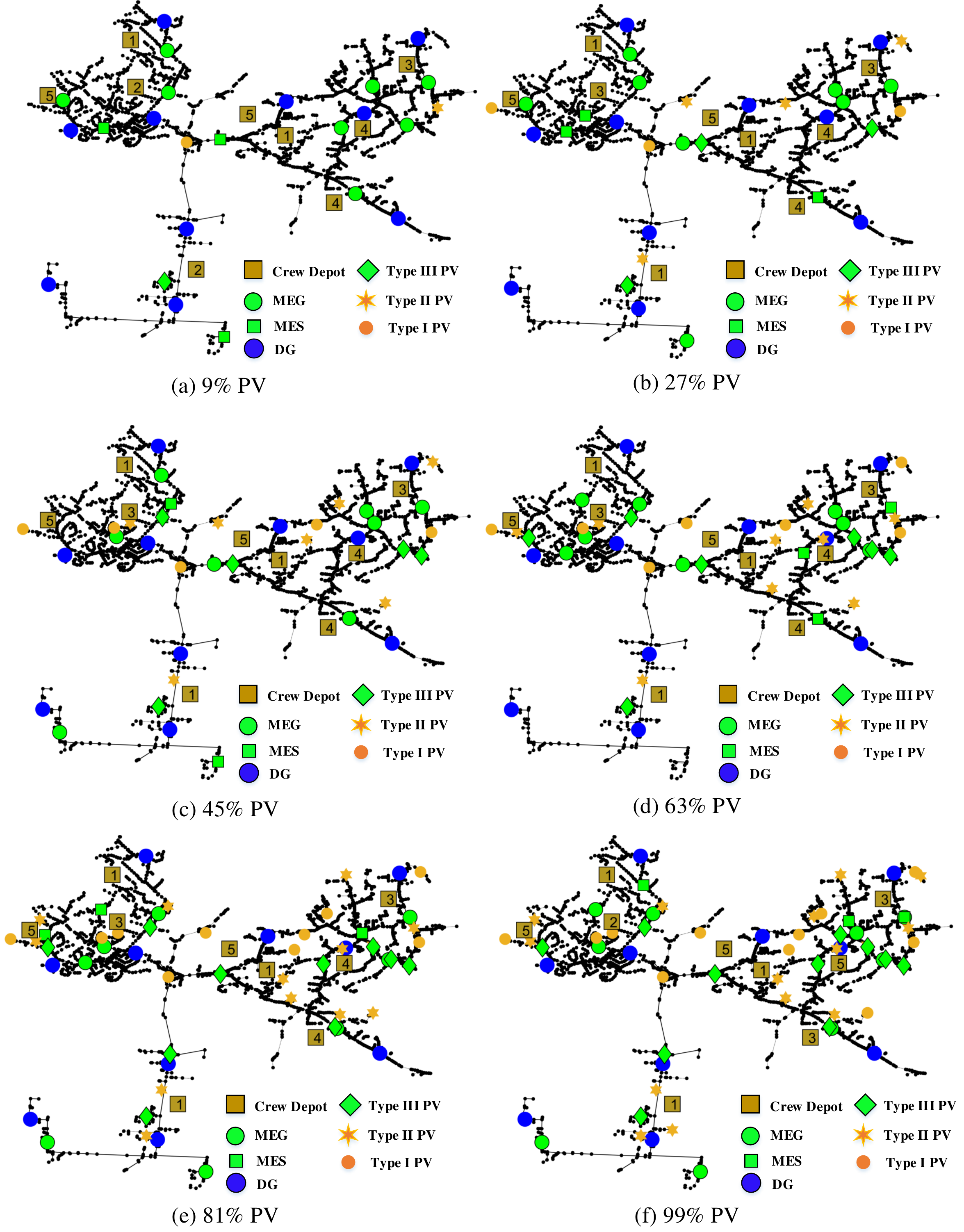}
\caption{Pre-event resource allocation results with different PV penetration levels.}
\label{large_diff_PVs}
\end{figure}

Based on the results of Figure. \ref{large_diff_PVs}, it can be observed that different PV penetration levels have different allocation results of the flexible resources, including the positions of MEGs, MESs and number of repair crews.  

\begin{figure}[H]
\centering\includegraphics[width=1.0\linewidth]{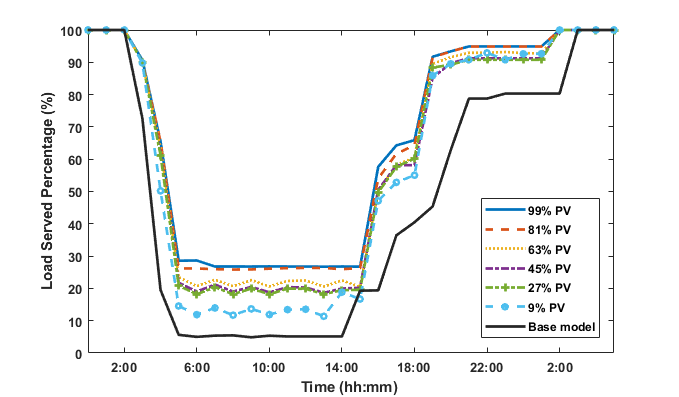}
\caption{Load served percentage comparison of proposed model with various PV penetration level and base model.}
\label{load_diff_PVs}
\end{figure}

Figure. \ref{load_diff_PVs} shows the percentage of power served during the event, and after the repair process starts. Table \ref{resilience} and Table \ref{outage} compare the amount of load served and average outage duration with different levels of PV penetration.

Based on the results from Figure. \ref{load_diff_PVs}, Table \ref{resilience} and Table \ref{outage}, it can be seen that the penetration of PV contributes to enhancing system resilience. Approximately 31.13\% more loads are served compared to the base model when the proposed method with 99\% PV penetration is used. Also, the average outage duration decreased by 31.12\%. However, compared with the case of 81\% PV penetration level, the proposed method with 99\% PV penetration does not have significant improvement. 
\begin{table}[H]
		\centering
		\renewcommand{\arraystretch}{1.3}		
		\caption{The amount of load served and resilience improvement with different level of PV penetration}
	    \label{resilience}
\begin{tabular}{lccll}
\hline
\textbf{\begin{tabular}[c]{@{}l@{}}PV Penetration \\ Percentage\end{tabular}} & \multicolumn{1}{l}{\textbf{\begin{tabular}[c]{@{}l@{}}Load Served \\ (kWh)\end{tabular}}} & \multicolumn{1}{l}{\textbf{\begin{tabular}[c]{@{}l@{}}Resilience Improvement \\ Percentage(\%)\end{tabular}}} \\\hline
\textbf{0}                                                               & 251,210.72                                                                                & -                                                                                                              \\
\textbf{9\%}                                                               & 318,668.37                                                                                & 26.85                                                                                                              \\
\textbf{27\%}                                                               & 335,525.77                                                                                & 33.56                                                                                                              \\
\textbf{45\%}                                                               & 336,710.74                                                                                & 34.04                                                                                                              \\
\textbf{63\%}                                                               & 344,588.22                                                                                & 37.17                                                                                                              \\
\textbf{81\%}                                                         & 360,668.04                                                                                & 43.57                                                                                                           \\
\textbf{99\%}                                                            & 364,785.93                                                                                & 45.21                \\\hline                                                                                          
\end{tabular}
\end{table}

\begin{table}[H]
		\centering
		\renewcommand{\arraystretch}{1.3}		
		\caption{The amount of average outage duration and outage decreased percentage with different level of PV penetration}
	    \label{outage}
\begin{tabular}{lccll}
\hline
\textbf{\begin{tabular}[c]{@{}l@{}}PV Penetration \\ Percentage\end{tabular}} & \multicolumn{1}{l}{\textbf{\begin{tabular}[c]{@{}l@{}}Average Outage \\ Duration (hour)\end{tabular}}} & \multicolumn{1}{l}{\textbf{\begin{tabular}[c]{@{}l@{}}Outage Decreased \\ Percentage(\%)\end{tabular}}} \\\hline
\textbf{0}                                                               & 14.69                                                                                & -                                                                                                              \\
\textbf{9\%}                                                               & 12.33                                                                                & 16.07                                                                                                              \\
\textbf{27\%}                                                               & 11.72                                                                                & 20.22                                                                                                              \\
\textbf{45\%}                                                               & 11.65                                                                                & 20.69                                                                                                              \\
\textbf{63\%}                                                               & 11.21                                                                                & 23.69                                                                                                              \\
\textbf{81\%}                                                         & 10.45                                                                                & 28.86                                                                                                           \\
\textbf{99\%}                                                            & 10.12                                                                                & 31.11                \\\hline                                                                                          
\end{tabular}
\end{table}

\section{Conclusion}\label{sec:Con}
In this paper, we develop a two-stage stochastic pre-event resource allocation method for upcoming extreme events, which enables faster and more efficient post-event restoration. The proposed pre-event method leverages the pre-allocation of mobile resources, fuel resources and labor resources. It also facilitates the benefits of distributed PV systems in resilience improvement of distribution grids. According to the case studies, we have the following observations: 
(i) Compared to the base model without pre-event resource allocation, the proposed pre-event preparation model can serve more loads and reduce the outage duration. 
(ii) Based on the response of the system with different PV penetration levels, it can be observed that the proposed pre-event preparation model with high PV penetration can further improve system resilience and reduce the outage duration. Therefore, PV systems can play a critical role in improving distribution grid resilience and further promote the renewable energy deployment.  
(iii) By considering the trade-off between solution accuracy and computation efficiency, the result of MRP indicates that the proposed model’s solutions with a limited number of scenarios can be very stable and of high quality. The scalability of the proposed pre-event preparation model is verified with a large-scale system. 

\section*{Acknowledgement}\label{sec:ack}
This work was supported  by the U.S. Department of Energy Wind Energy Technologies Office under Grant DE-EE0008956.





\bibliographystyle{elsarticle-num-names}
\bibliography{sample.bib}

\begin{thebibliography}{26}
\expandafter\ifx\csname natexlab\endcsname\relax\def\natexlab#1{#1}\fi
\providecommand{\url}[1]{\texttt{#1}}
\providecommand{\href}[2]{#2}
\providecommand{\path}[1]{#1}
\providecommand{\DOIprefix}{doi:}
\providecommand{\ArXivprefix}{arXiv:}
\providecommand{\URLprefix}{URL: }
\providecommand{\Pubmedprefix}{pmid:}
\providecommand{\doi}[1]{\href{http://dx.doi.org/#1}{\path{#1}}}
\providecommand{\Pubmed}[1]{\href{pmid:#1}{\path{#1}}}
\providecommand{\bibinfo}[2]{#2}
\ifx\xfnm\relax \def\xfnm[#1]{\unskip,\space#1}\fi
\bibitem[{Salman et~al.(2015)Salman, Li, and Stewart}]{salman2015evaluating}
\bibinfo{author}{A.~M. Salman}, \bibinfo{author}{Y.~Li}, \bibinfo{author}{M.~G.
  Stewart},
\newblock \bibinfo{title}{Evaluating system reliability and targeted hardening
  strategies of power distribution systems subjected to hurricanes},
\newblock \bibinfo{journal}{Reliability Engineering \& System Safety}
  \bibinfo{volume}{144} (\bibinfo{year}{2015}) \bibinfo{pages}{319--333}.
\bibitem[{of~the President(2020)}]{WH_report}
\bibinfo{author}{E.~O. of~the President}, \bibinfo{title}{Economic benefits of
  increasing electric grid resilience to weather outages},
  \bibinfo{type}{Technical Report}, White House Tech. Rep.,
  \bibinfo{year}{2020}.
\bibitem[{{Gholami} et~al.(2019){Gholami}, {Shekari}, and {Grijalva}}]{pre_3}
\bibinfo{author}{A.~{Gholami}}, \bibinfo{author}{T.~{Shekari}},
  \bibinfo{author}{S.~{Grijalva}},
\newblock \bibinfo{title}{Proactive management of microgrids for resiliency
  enhancement: An adaptive robust approach},
\newblock \bibinfo{journal}{IEEE Trans. Sustain. Energy} \bibinfo{volume}{10}
  (\bibinfo{year}{2019}) \bibinfo{pages}{470--480}.
\bibitem[{{Wang} et~al.(2017){Wang}, {Hou}, {Qiu}, {Lei}, and {Liu}}]{pre_4}
\bibinfo{author}{C.~{Wang}}, \bibinfo{author}{Y.~{Hou}},
  \bibinfo{author}{F.~{Qiu}}, \bibinfo{author}{S.~{Lei}},
  \bibinfo{author}{K.~{Liu}},
\newblock \bibinfo{title}{Resilience enhancement with sequentially proactive
  operation strategies},
\newblock \bibinfo{journal}{IEEE Trans. Power Syst.} \bibinfo{volume}{32}
  (\bibinfo{year}{2017}) \bibinfo{pages}{2847--2857}.
\bibitem[{{Panteli} et~al.(2017){Panteli}, {Mancarella}, {Trakas},
  {Kyriakides}, and {Hatziargyriou}}]{pre_7}
\bibinfo{author}{M.~{Panteli}}, \bibinfo{author}{P.~{Mancarella}},
  \bibinfo{author}{D.~N. {Trakas}}, \bibinfo{author}{E.~{Kyriakides}},
  \bibinfo{author}{N.~D. {Hatziargyriou}},
\newblock \bibinfo{title}{Metrics and quantification of operational and
  infrastructure resilience in power systems},
\newblock \bibinfo{journal}{IEEE Trans. Power Syst.} \bibinfo{volume}{32}
  (\bibinfo{year}{2017}) \bibinfo{pages}{4732--4742}.
\bibitem[{{Wang} et~al.(2004){Wang}, {Sarker}, {Mann}, and
  {Triantaphyllou}}]{pre_9}
\bibinfo{author}{S.~{Wang}}, \bibinfo{author}{B.~R. {Sarker}},
  \bibinfo{author}{L.~{Mann}}, \bibinfo{author}{E.~{Triantaphyllou}},
\newblock \bibinfo{title}{Resource planning and a depot location for electric
  power restoration},
\newblock \bibinfo{journal}{Euro. J. Oper. Res.} \bibinfo{volume}{155}
  (\bibinfo{year}{2004}) \bibinfo{pages}{22--43}.
\bibitem[{{Arif} et~al.(2018){Arif}, {Wang}, {Wang}, and {Chen}}]{B_1}
\bibinfo{author}{A.~{Arif}}, \bibinfo{author}{Z.~{Wang}},
  \bibinfo{author}{J.~{Wang}}, \bibinfo{author}{C.~{Chen}},
\newblock \bibinfo{title}{Power distribution system outage management with
  co-optimization of repairs, reconfiguration, and {DG} dispatch},
\newblock \bibinfo{journal}{IEEE Trans. Smart Grid} \bibinfo{volume}{9}
  (\bibinfo{year}{2018}) \bibinfo{pages}{4109--4118}.
\bibitem[{{Arif} et~al.(2020){Arif}, {Wang}, {Chen}, and {Chen}}]{B_3}
\bibinfo{author}{A.~{Arif}}, \bibinfo{author}{Z.~{Wang}},
  \bibinfo{author}{C.~{Chen}}, \bibinfo{author}{B.~{Chen}},
\newblock \bibinfo{title}{A stochastic multi-commodity logistic model for
  disaster preparation in distribution systems},
\newblock \bibinfo{journal}{IEEE Trans. Smart Grid} \bibinfo{volume}{11}
  (\bibinfo{year}{2020}) \bibinfo{pages}{565--576}.
\bibitem[{{Taheri} et~al.(2019){Taheri}, {Safdarian}, {Moeini-Aghtaie}, and
  {Lehtonen}}]{pre_1}
\bibinfo{author}{B.~{Taheri}}, \bibinfo{author}{A.~{Safdarian}},
  \bibinfo{author}{M.~{Moeini-Aghtaie}}, \bibinfo{author}{M.~{Lehtonen}},
\newblock \bibinfo{title}{Enhancing resilience level of power distribution
  systems using proactive operational actions},
\newblock \bibinfo{journal}{IEEE Access} \bibinfo{volume}{7}
  (\bibinfo{year}{2019}) \bibinfo{pages}{137378--137389}.
\bibitem[{{Lei} et~al.(2018){Lei}, {Wang}, {Chen}, and {Hou}}]{pre_5}
\bibinfo{author}{S.~{Lei}}, \bibinfo{author}{J.~{Wang}},
  \bibinfo{author}{C.~{Chen}}, \bibinfo{author}{Y.~{Hou}},
\newblock \bibinfo{title}{Mobile emergency generator pre-positioning and
  real-time allocation for resilient response to natural disasters},
\newblock \bibinfo{journal}{IEEE Trans. Smart Grid} \bibinfo{volume}{9}
  (\bibinfo{year}{2018}) \bibinfo{pages}{2030--2041}.
\bibitem[{{Lei} et~al.(2019){Lei}, {Chen}, {Zhou}, and {Hou}}]{pre_6}
\bibinfo{author}{S.~{Lei}}, \bibinfo{author}{C.~{Chen}},
  \bibinfo{author}{H.~{Zhou}}, \bibinfo{author}{Y.~{Hou}},
\newblock \bibinfo{title}{Routing and scheduling of mobile power sources for
  distribution system resilience enhancement},
\newblock \bibinfo{journal}{IEEE Trans. Smart Grid} \bibinfo{volume}{10}
  (\bibinfo{year}{2019}) \bibinfo{pages}{5650--5662}.
\bibitem[{{Kim} and {Dvorkin}(2019)}]{pre_2}
\bibinfo{author}{J.~{Kim}}, \bibinfo{author}{Y.~{Dvorkin}},
\newblock \bibinfo{title}{Enhancing distribution system resilience with mobile
  energy storage and microgrids},
\newblock \bibinfo{journal}{IEEE Trans. Smart Grid} \bibinfo{volume}{10}
  (\bibinfo{year}{2019}) \bibinfo{pages}{4996--5006}.
\bibitem[{{Samara} et~al.(2020){Samara}, {Shaaban}, and {Osman}}]{pre_10}
\bibinfo{author}{S.~{Samara}}, \bibinfo{author}{M.~F. {Shaaban}},
  \bibinfo{author}{A.~H. {Osman}},
\newblock \bibinfo{title}{Optimal management of mobile energy generation and
  storage systems},
\newblock \bibinfo{journal}{IEEE Access} \bibinfo{volume}{8}
  (\bibinfo{year}{2020}) \bibinfo{pages}{203890--203900}.
\bibitem[{Belding et~al.(2020)Belding, Walker, and Watson}]{NREL_report}
\bibinfo{author}{S.~Belding}, \bibinfo{author}{A.~Walker},
  \bibinfo{author}{A.~Watson}, \bibinfo{title}{Will solar panels help when the
  power goes out?}, \bibinfo{type}{Technical Report}, National Renewable Energy
  Tech. Rep., \bibinfo{year}{2020}.
\bibitem[{Rockafellar and Wets(1991)}]{T2_7}
\bibinfo{author}{R.~T. Rockafellar}, \bibinfo{author}{R.~J.-B. Wets},
\newblock \bibinfo{title}{Scenarios and policy aggregation in optimization
  under uncertainty},
\newblock \bibinfo{journal}{Mathematics of operations research}
  \bibinfo{volume}{16} (\bibinfo{year}{1991}) \bibinfo{pages}{119--147}.
\bibitem[{Watson and Woodruff(2011)}]{T2_8}
\bibinfo{author}{J.-P. Watson}, \bibinfo{author}{D.~L. Woodruff},
\newblock \bibinfo{title}{Progressive hedging innovations for a class of
  stochastic mixed-integer resource allocation problems},
\newblock \bibinfo{journal}{Computational Management Science}
  \bibinfo{volume}{8} (\bibinfo{year}{2011}) \bibinfo{pages}{355--370}.
\bibitem[{{Ma} et~al.(2018){Ma}, {Chen}, and {Wang}}]{scen_ma}
\bibinfo{author}{S.~{Ma}}, \bibinfo{author}{B.~{Chen}},
  \bibinfo{author}{Z.~{Wang}},
\newblock \bibinfo{title}{Resilience enhancement strategy for distribution
  systems under extreme weather events},
\newblock \bibinfo{journal}{IEEE Trans. Smart Grid} \bibinfo{volume}{9}
  (\bibinfo{year}{2018}) \bibinfo{pages}{1442--1451}.
\bibitem[{{Ma} et~al.(2019){Ma}, {Li}, {Wang}, and {Qiu}}]{B_4}
\bibinfo{author}{S.~{Ma}}, \bibinfo{author}{S.~{Li}},
  \bibinfo{author}{Z.~{Wang}}, \bibinfo{author}{F.~{Qiu}},
\newblock \bibinfo{title}{Resilience-oriented design of distribution systems},
\newblock \bibinfo{journal}{IEEE Trans. Power Syst.} \bibinfo{volume}{34}
  (\bibinfo{year}{2019}) \bibinfo{pages}{2880--2891}.
\bibitem[{{Chen} et~al.(2018){Chen}, {Chen}, {Wang}, and {Butler-Purry}}]{T2_1}
\bibinfo{author}{B.~{Chen}}, \bibinfo{author}{C.~{Chen}},
  \bibinfo{author}{J.~{Wang}}, \bibinfo{author}{K.~L. {Butler-Purry}},
\newblock \bibinfo{title}{Sequential service restoration for unbalanced
  distribution systems and microgrids},
\newblock \bibinfo{journal}{IEEE Trans. Power Syst.} \bibinfo{volume}{33}
  (\bibinfo{year}{2018}) \bibinfo{pages}{1507--1520}.
\bibitem[{{Arif} et~al.(2020){Arif}, {Wang}, {Chen}, and {Wang}}]{T2_2}
\bibinfo{author}{A.~{Arif}}, \bibinfo{author}{Z.~{Wang}},
  \bibinfo{author}{C.~{Chen}}, \bibinfo{author}{J.~{Wang}},
\newblock \bibinfo{title}{Repair and resource scheduling in unbalanced
  distribution systems using neighborhood search},
\newblock \bibinfo{journal}{IEEE Trans. Smart Grid} \bibinfo{volume}{11}
  (\bibinfo{year}{2020}) \bibinfo{pages}{673--685}.
\bibitem[{{Melhem} et~al.(2018){Melhem}, {Grunder}, {Hammoudan}, and
  {Moubayed}}]{T2_6}
\bibinfo{author}{F.~Y. {Melhem}}, \bibinfo{author}{O.~{Grunder}},
  \bibinfo{author}{Z.~{Hammoudan}}, \bibinfo{author}{N.~{Moubayed}},
\newblock \bibinfo{title}{Energy management in electrical smart grid
  environment using robust optimization algorithm},
\newblock \bibinfo{journal}{IEEE Trans. Industry Applications}
  \bibinfo{volume}{54} (\bibinfo{year}{2018}) \bibinfo{pages}{2714--2726}.
\bibitem[{{Zhang} et~al.(2019){Zhang}, {Dehghanpour}, and {Wang}}]{QZ_CVR}
\bibinfo{author}{Q.~{Zhang}}, \bibinfo{author}{K.~{Dehghanpour}},
  \bibinfo{author}{Z.~{Wang}},
\newblock \bibinfo{title}{Distributed {CVR} in unbalanced distribution systems
  with {PV} penetration},
\newblock \bibinfo{journal}{IEEE Trans. Smart Grid} \bibinfo{volume}{10}
  (\bibinfo{year}{2019}) \bibinfo{pages}{5308--5319}.
\bibitem[{EPRI(2019)}]{Opendss}
\bibinfo{author}{EPRI}, \bibinfo{title}{{OPENDSS} test circuits},
  \bibinfo{year}{Apr.2019}. \URLprefix
  \url{https://sourceforge.net/p/electricdss/discussion/beginners.html}.
\bibitem[{{Arritt} and {Dugan}(2010)}]{8500}
\bibinfo{author}{R.~F. {Arritt}}, \bibinfo{author}{R.~C. {Dugan}},
\newblock \bibinfo{title}{The {IEEE} 8500-node test feeder},
\newblock in: \bibinfo{booktitle}{IEEE PES T\&D Conference},
  \bibinfo{year}{2010}, pp. \bibinfo{pages}{1--6}.
\bibitem[{Ma(2020)}]{ma2020resilience}
\bibinfo{author}{S.~Ma},
\newblock \bibinfo{title}{Resilience-oriented design and proactive preparedness
  of electrical distribution system},
\newblock \bibinfo{journal}{PhD Thesis}  (\bibinfo{year}{2020}).
\bibitem[{Hart et~al.(2017)Hart, Laird, Watson, Woodruff, Hackebeil, Nicholson,
  and Siirola}]{T2_11}
\bibinfo{author}{W.~E. Hart}, \bibinfo{author}{C.~D. Laird},
  \bibinfo{author}{J.-P. Watson}, \bibinfo{author}{D.~L. Woodruff},
  \bibinfo{author}{G.~A. Hackebeil}, \bibinfo{author}{B.~L. Nicholson},
  \bibinfo{author}{J.~D. Siirola}, \bibinfo{title}{Pyomo-optimization modeling
  in python}, volume~\bibinfo{volume}{67}, \bibinfo{publisher}{Springer},
  \bibinfo{year}{2017}.

\end{thebibliography}







\end{document}